\documentclass[twocolumn,aps,prl,showpacs,superscriptaddress,longbibliography,10pt]{revtex4-1} %

\usepackage[colorlinks=true,linkcolor=blue,citecolor=blue]{hyperref}
\usepackage{amsfonts}
\usepackage{subfigure}
\usepackage{amsmath}
\usepackage{txfonts}
\usepackage{amssymb}
\usepackage{amsbsy}
\usepackage{epsfig}
\usepackage{graphicx}
\usepackage{epstopdf}
\usepackage{mathdots}
\usepackage{color}
\usepackage{cleveref}

\begin{document}
\title{Universal characteristics of one-dimensional non-Hermitian superconductors}

\author{Yang Cao}
\affiliation{Department of Physics, Jiangsu University, Zhenjiang, 212013, China}

\author{Yang Li}
\affiliation{Department of Physics, Jiangsu University, Zhenjiang, 212013, China}

\author{Yuanping Chen} \altaffiliation{chenyp@ujs.edu.cn}
\affiliation{Department of Physics, Jiangsu University, Zhenjiang, 212013, China}

\author{Xiaosen Yang} \altaffiliation{yangxs@ujs.edu.cn}
\affiliation{Department of Physics, Jiangsu University, Zhenjiang, 212013, China}

\date{\today}

\begin{abstract}
We establish a non-Bloch band theory for one-dimensional(1D) non-Hermitian topological superconductors. The universal physical properties of non-Hermitian topological superconductors are revealed based on the theory. According to the particle-hole symmetry, there exist reciprocal particle and hole loops of generalized Brillouin zone (GBZ). The critical point of quantum phase transition, where the energy gap closes, appears when the particle and hole loops intersect and their values of GBZ satisfy $|\beta|=1$. If the non-Hermitian system has skin modes, these modes should be $Z_{2}$ style, i.e., the corresponding eigenstates of particle and hole localize at opposite ends of an open chain, respectively. The non-Bloch band theory is applied to two examples, non-Hermitian $p$- and $s$-wave topological superconductors. Topological phase transitions occur at $\beta_{c}=\pm 1$ in the two systems. In terms of Majorana Pfaffian, a $Z_{2}$ non-Bloch topological invariant is defined to establish the non-Hermitian bulk-boundary correspondence in non-Hermitian superconductors.
\end{abstract}

\maketitle

{\it Introduction.--}
Non-Hermitian systems\cite{benderPhysRevLett1998, Diehlnaturephys2011, huicaoRevModPhys2015, zhenbonature2015, PhysRevLettLee2016, HosseinNature2017, zhongwangPhysRevLett2018a, GanainyNaturePhysics2018, LonghiPRL2019, zhaihuiNP2020}, described by non-Hermitian Hamiltonians, exhibit a new physical world beyond Hermitian systems from classic\cite{EzawaPRB2019, YoshidaAPS2020, HofmannPRR2020, HelbigNaturePhysics2020,WeiweiPhysRevLett2018, Gao2020arXiv} to quantum \cite{AloisNature2012,lunaturepho2014,XiaoLeiNaturePhysics2020,liarxiv2020,ShenPhysRevLett2018b,GongPhysRevX2018, ShenPRL2018, ChenYuPhysRevB2018, KunstPhysRevLett2018, kouPhysRevB2020, HaoranPhysRevLett2020,YoungwoonPhysRevLett2010, leePhysRevX2014, XuPhysRevLett2017, LinhuPhysRevLett2020,ZengPRA2017, MenkePhysRevB2017, Shibata2019PhRvB, Terrier2020PhRvR} in recent years. For example, in a non-Hermitian topological insulator, the eigenstates localize at the boundaries under open-boundary condition(OBC)  rather than extend over the bulk\cite{zhongwangPhysRevLett2018a, zhongwangPhysRevLett2018b, LonghiPhysRevResearch2019, chenshuPRB2019, JiangbinhybridPRL2019,  liuPhysRevLett2019, PhysRevLettwangzhong2019a, LuoPRL2019, li2020critical, haga2020liouvillian, yi2020nonhermitian, OkumaPhysRevLett2020,zengPRR2020, liu2020helical,WanPhysRevBsecond,wangNBPT2021,FNHPRBcao}. This extraordinary skin effect confirmed by several experiments\cite{ghatak2019observation,XiaoLeiNaturePhysics2020,HelbigNaturePhysics2020,  HofmannPRR2020, xuepengNBPT2020} has never been found in the Hermitian systems, and it cannot be explained by the conventional bulk-boundary correspondence (BBC) in the framework of Bloch theory. To reveal the novel BBC, a non-Bloch band theory for non-Hermitian topological insulator is introduced, in which the concept of Brillouin zone(BZ) in the Bloch theory is extended to GBZ\cite{zhongwangPhysRevLett2018a, zhongwangPhysRevLett2018b, YokomizoPhysRevLett2019,weiyiPhysRevB2019, PhysRevLettwangzhong2019b,  KawabataPRX2019, yangarxiv2019,xuearxiv2020, zhangPRLcorrespondence, LonghiPhysRevLett2020,FNHPRBcao}. Correspondingly, non-Bloch topological invariants can be defined in the GBZ\cite{zhongwangPhysRevLett2018a, zhongwangPhysRevLett2018b, YokomizoPhysRevLett2019}, which describe fingerprints of the non-Hermitian topological insulators like the topological invariants for the Hermitian topological insulators.

Following the studies on non-Hermitian topological insulator, non-Hermitian topological superconductor has attracted much attention very recently\cite{WangPhysRevA2015,KawabataPhysRevB2018B, GhatakPhysRevB2018, Avila2019CmPhy, Kawabataurecomm2019, OkugawaPhysRevB2019}. Topological superconductor is a cousin of topological insulator in the topological family. In a Hermitian topological superconductor, there exist Majorana zero modes(MZMs) at topological defects such as boundaries, vortices or domain walls\cite{ReadPhysRevB2000, KitaevPhyU2001, IvanovPhysRevLett2001, FuPhysRevLett2008, Sternnature2010, MourikScience2012}. The robust MZMs hold the promising applications in fault-tolerant quantum computing\cite{NayakRevModPhys2008, Aliceanature2011, LianNAS}. A few preliminary researches indicate that the non-Hermitian superconductors indeed have exotic phenomena, such as robust MZMs\cite{ZengPhysRevA2016, Li2017arXiv, McDonaldPhysRevX2018, LieuPhysRevB2019, OkumaPRL2019, ZhouPhysRevB2020, YangPhysRevA2020, Zhao2020arXiv} and non-Hermitian skin effect\cite{Liu2020arXiv}. However, the universal physical characteristics of the non-Hermitian superconductor systems have not been uncovered far away, because a proper non-Bloch band theory for the non-Hermitian systems has not been established yet.

In this letter, we establish a non-Bloch band theory for 1D non-Hermitian superconductors and reveal the universal properties of the systems. Because of the particle-hole symmetry(PHS), the GBZ of the systems are determined as paired reciprocal loops. Remarkably, the non-Hermitian skin effect of the superconductors is a $Z_{2}$ skin effect. The energy gap of a non-Hermitian superconductor under OBC closes only when the paired loops of GBZ intersect at critical values with $|\beta_{c}|=1$. These universal properties are clarified in two simple non-Hermitian $p$- and $s$-wave superconductor systems. The GBZ of the two systems are one and two paired reciprocal loops, respectively. A $Z_{2}$ skin effect is observed in the two non-Hermitian systems, where the corresponding eigenstates of particle and hole localize at opposite ends of an open chain. Topological phase transitions occur at critical points of the GBZ with $\beta_{c}=\pm1$. In addition, a $Z_{2}$ non-Bloch topological invariant is defined in terms of the Pfaffian of the Majorana representation matrix, and the non-Hermitian BBC of the non-Hermitian superconductors are established.

\begin{figure}
\includegraphics[width=0.48\textwidth]{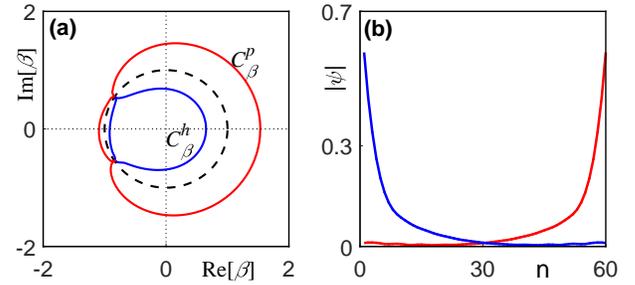}
\caption{(a) The GBZ of non-Hermitian superconductors is paired reciprocal loops of particle $C_{\beta}^{p}$ and hole $C_{\beta}^{h}$ with $\beta_{p} = \beta^{-1}_{h}$. (b) The corresponding particle (red) and hole (blue) eigenstates with $|\beta(\pm E)|\neq 1$ localize at opposite boundaries of an open chain.} \label{gbze}
\end{figure}

{\it Non-Bloch band theory and $Z_{2}$ skin effect.--} For a Hermitian superconductor, the periodic-boundary condition(PBC) energies of the Bloch Hamiltonian $H(k)$ with real-valued wave numbers $k$ are asymptotically identical to the energies of the system under OBC. The Majorana edge modes of a topological phase can be characterized by a topological invariant of the periodic system, which is named bulk-boundary correspondence. Therefore, the Bloch Hamiltonian captures the properties of the system under OBC. In contrast to Hermitian case, the Bloch Hamiltonian of a non-Hermitian superconductor no longer always give the properties of the system under OBC, such as the spectra and non-Hermitian skin effect. To understand the properties of the open system, it is necessary to go beyond the Bloch band theory by generalizing the real-valued wave numbers $k$ to a complex plane. Then, a non-Bloch Hamiltonian $H(\beta)$ can be obtained by generalization of the Bloch Hamiltonian with $\beta:=e^{ik}$. Here, $\beta$ is extended from a unit circle in Hermitian superconductors to the entire complex plane in non-Hermitian superconductors. We can determined $\beta$ with the complex eigenvalues by the characteristic equation:
\begin{eqnarray}
f(\beta,E)=\det [H(\beta) - E] = 0,
\label{eigeneq}
\end{eqnarray}
which is an even-order algebraic equation in terms of $\beta$, in generally. We number the solutions $\beta_{i}$($i=1,2,...,2M$) of the eigenvalue equation so as to satisfy $|\beta_{1}|\leq |\beta_{2}|\leq ...\leq |\beta_{2M}|$. Then, the continuum bands condition is given by\cite{zhongwangPhysRevLett2018a, YokomizoPhysRevLett2019, zhangPRLcorrespondence}:
\begin{eqnarray}
|\beta_{M}| = |\beta_{M+1}|,
\label{condition}
\end{eqnarray}
and the trajectories of $\beta_{M}$ and $\beta_{M+1}$ give the GBZ $C_{\beta}$. In the conception of the GBZ, the non-Bloch Hamiltonian $H(\beta)$ gives the spectra and the eigenstates of the open system.
For a non-Hermitian superconductor, the non-Bloch Hamiltonian $H(\beta)$ preserves the PHS:
\begin{eqnarray}
\mathcal{C} H^{T} (\beta) \mathcal{C}^{-1} = - H (\beta^{-1}),
\label{phs}
\end{eqnarray}
with $\mathcal{C}^{2} =\pm 1$. Thus, when $\beta$ is a solution of the Eq.(\ref{eigeneq}) for a eigenvalue $E$, we have
\begin{eqnarray}
\det \left[ H(\beta) - E \right] &=& \det \left[ - \mathcal{C}  H^{T}(\beta^{-1}) \mathcal{C}^{-1} - E \right] \nonumber\\
&=& \det \left[  H(\beta^{-1}) + E \right] = 0,
\label{eequation}
\end{eqnarray}
which implies that $\beta^{-1}$ is a solution for a eigenvalue $-E$. Then, the $\beta$ of the GBZ must satisfy
\begin{eqnarray}
\beta(E) = \beta^{-1}(-E),
\label{brecip}
\end{eqnarray}
which indicate that the GBZ of a non-Hermitian superconductor are paired particle and hole loops.
Figure \ref{gbze}(a) shows the typical GBZ of a two band non-Hermitian superconductor. The corresponding particle($C_{\beta}^{p}$) and hole($C_{\beta}^{h}$) loops are reciprocal to each other with $\beta_{p}=\beta_{h}^{-1}$.

\begin{figure*}
\centering
\includegraphics[width=0.95\textwidth]{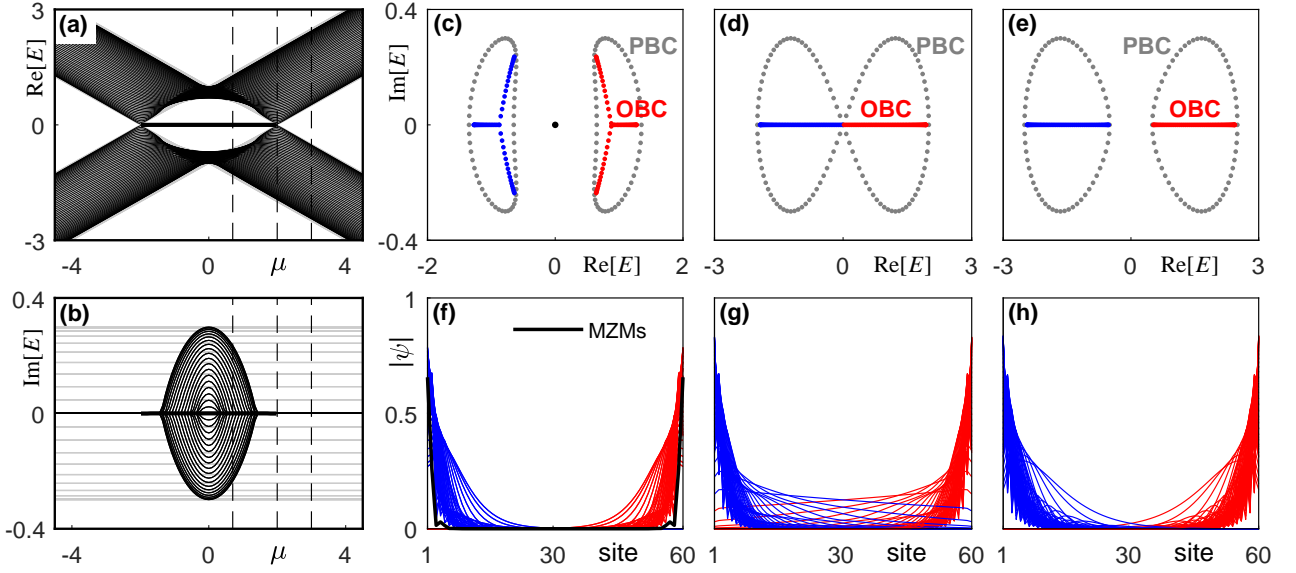}
\caption{(a) Real and (b) imaginary parts of energy spectra in the non-Hermitian Kitaev chain under open boundaries condition (black) and  periodic boundaries condition (gray) with lattice size $L = 60$. The bulk energy spectra of the two cases are significant different.  The robust Majorana zero modes (black lines) appear in the topological phase for region $\mu \in [-2, 2]$. (c)-(e) The energy spectra in an open chain (red and blue) and in a periodic chain (gray). The values of parameters are $t_{L} = 1.3$, $t_{R} = 0.7$ and $\Delta = 0.7$ with (c) $\mu = 0.7$, (d) $\mu = 2.0$ and (e) $\mu = 3.0$, respectively. (f)-(h) The eigenstates of the Kitaev chain under open boundaries condition for (c)-(e).}\label{skin}
\end{figure*}

For a Hermitian superconductor, the phase transition occurs when the energy gap between particle and hole closes at critical real-valued wave numbers $k_{c}$ with $E(k_{c})=0$.  But the particle loops and the hole loops of GBZ are not always coincide for a non-Hermitian superconductor. Is there any universal features for the gap closing of the non-Hermitian superconductors? Here, we show how the energy gap closes under OBC.
When the energy gap closes with $E=0$, both of $\beta_{c,i}$ and $\beta_{c,i}^{-1}$ are the solutions of characteristic equation $f(\beta_{c},E=0)=0$ without changing the order in terms of $\beta_{c}$. Thus, the corresponding solutions $\beta_{c,i}$ satisfy $\beta_{c,i} = \beta_{c,2M-i+1}^{-1}$. On account of the GBZ condition (Eq.\ref{condition}), we obtain the gap closing condition:
\begin{eqnarray}
|\beta_{c}| = 1.
\label{gapcondition}
\end{eqnarray}
Figure \ref{gbze}(a) shows an example of two gap closing points of GBZ. The particle($C_{\beta}^{p}$) and hole($C_{\beta}^{h}$) loops of GBZ intersect at unit circle with $|\beta|=1$ where gap closes with $E(\beta)=0$.

For a non-Hermitian superconductor, a right eigenvector can be defined by $H(\beta(E))|\beta(E), E\rangle=E|\beta(E), E\rangle$. According to Eq.\ref{brecip}, a particle eigenvector $|\beta(E), E\rangle$ can be mapped to a hole eigenvector $|\beta(-E), -E\rangle$ under the particle-hole transformation with
\begin{eqnarray}
|\beta(E), E\rangle = \mathcal{C} |\beta(-E), -E\rangle.
\label{phtransf}
\end{eqnarray}
This indicate that the corresponding eigenstates of particle and hole are localized at the opposite boundaries of the system under OBC when $|\beta(E)|\neq 1$\cite{zhongwangPhysRevLett2018a, OkumaPhysRevLett2020, zhangPRLcorrespondence} as shown in Fig.\ref{gbze}(b). Therefore, the non-Hermitian skin modes of a superconductor are $Z_{2}$ skin modes, which is protected by the PHS.

Due to the above non-Hermitian features, a non-Hermitian superconductor under OBC is no longer captured by a Bloch Hamiltonian, but by a non-Bloch Hamiltonian. Then, a topological invariant also should be defined on the GBZ to characterize the topological properties of the system. To be concrete, we perform the non-Bloch band theory into two typical systems: non-Hermitian $p$- and $s$-wave superconductors. With the help of GBZ, a non-Bloch topological invariant is defined in terms of the Majorana Pfaffian to establish the non-Hermitian BBC for the non-Hermitian superconductors.

{\it Non-Hermitian Kitaev Model.--} The nonreciprocal Kitaev model describes a non-Hermitian spinless $p$-wave superconductor, which can be written in real-space as:
\begin{eqnarray}
H=\sum_{i}\left[-t_{L}c_{i}^{\dag}c_{i+1}-t_{R}c_{i+1}^{\dag}c_{i}+\Delta c_{i}c_{i+1}+\Delta c_{i+1}^{\dag}c_{i}^{\dag}-\mu c_{i}^{\dag}c_{i}\right],
\label{hamiltonian}
\end{eqnarray}
where $c_{i}^{\dag}(c_{i})$ is the creation (annihilation) operator on site $i$. $t_{L} (t_{R})$, $\Delta$ and $\mu$ are real parameters and denote the left (right) hopping amplitude, pairing amplitude and chemical potential respectively. When $t_{L} = t_{R}$ (Hermitian case), the OBC energies are asymptotically identical to the PBC energies. The MZMs locate at the ends of an open chain when $\mu \in [-2, 2]$\cite{KitaevPhyU2001}.

When $t_{L}\neq t_{R}$, the OBC energies will be different from PBC energies, especially in the imaginary parts.
Figures \ref{skin}(a)-(b) show the difference for the real and imaginary part, where the black(gray) curves represent the cases of OBC(PBC). The values of other parameters are $t_{L} = 1.3$, $t_{R} = 0.7$ and $\Delta = 0.7$. The imaginary part of the energies does not change with the chemical potential under PBC, but sharply collapse under OBC. Meanwhile, it is found that there are MZMs appearing in the open chain. Figures \ref{skin}(c)-(e) show three typical energy spectra with $\mu=0.7$, $2.0$ and $3.0$, respectively. The OBC energy spectra dramatically collapses compared with the PBC energy spectra. Figures \ref{skin}(f)-(h) show the corresponding eigenstates of an open Kitaev chain for Fig.\ref{skin}(c)-(e). We can find that the open chain exhibits a $Z_{2}$ skin effect. The eigenstates localize at the right end for particles and localize at the left end for holes. Meanwhile, the robust MZMs localize at both the right and the left ends as shown in Fig.\ref{skin}(f).

To exactly capture the non-Hermitian features, the Hamiltonian in Eq.\ref{hamiltonian} can be mapped to a non-Bloch Hamiltonian $H(\beta)$, expressed as:
\begin{eqnarray}
H(\beta) = h_{0}(\beta) + h_{y}(\beta) \sigma_{y} +h_{z}(\beta) \sigma_{z},
\label{hamilbeta}
\end{eqnarray}
here, $h_{0}(\beta) = \delta (\beta - \beta^{-1})$, $h_{y}(\beta) = i \Delta (\beta - \beta^{-1}) $ and $h_{z}(\beta) = \mu + t (\beta+ \beta^{-1})$ with $ 2t = t_{L} + t_{R}$ and  $2\delta = t_{L} - t_{R}$. The Hamiltonian $H(\beta)$ preserves the PHS: $\sigma_{x} H^{T}(\beta^{-1}) \sigma_{x} = - H(\beta)$, which ensures that the eigenvalues appear in pairs: $E_{\pm} (\beta) = h_{0}(\beta) \pm \sqrt{h_{y}^{2}(\beta) + h_{z}^{2}(\beta)}$.

\begin{figure}
\includegraphics[width=0.48\textwidth]{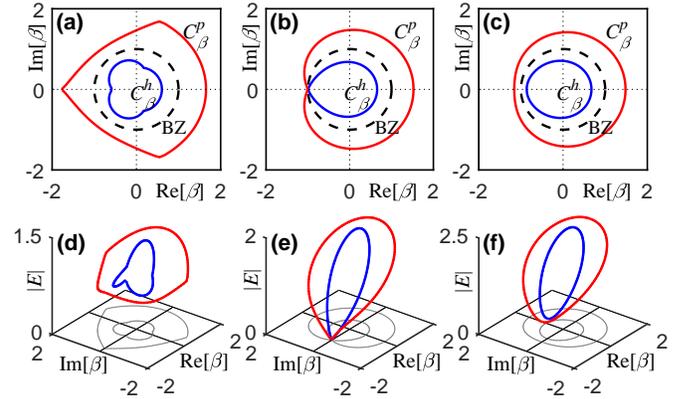}
\caption{(a)-(c) The GBZ (solid curves) and BZ (dashed curves) for Fig.\ref{skin}(b)-(d). The particle loop $C_{\beta}^{p}$ (red curve) and the hole loop $C_{\beta}^{h}$ (blue curve) are reciprocal to each other. (d)-(f) Absolute energy spectra of an open chain on the GBZ for (a)-(c).}\label{gbz}
\end{figure}

The characteristic equation of the Hamiltonian in Eq.\ref{hamilbeta} has four solutions. According to Eq.\ref{condition}, the GBZ of the non-Hermitian Kitaev model is given by the trajectories of $\beta_{2}$ and $\beta_{3}$ with $|\beta_{2}| = |\beta_{3}|$.
Figures \ref{gbz}(a)-(c) show the GBZ of the non-Hermitian Kitaev chain for Fig.\ref{skin}(c)-(e), respectively.  The loops of particles $C_{\beta}^{p}$ and holes $C_{\beta}^{h}$ are reciprocal to each other with $\beta_{p}=\beta_{h}^{-1}$. To reveal the gap closing, figures \ref{gbz}(d)-(f) show the absolute energy spectra in an open chain on the GBZ of Fig.\ref{gbz}(a)-(c). The inner loop of GBZ $C_{\beta}^{h}$ gives the energy spectra of holes with $\mathrm{Re}(E(\beta_{h})) < 0$, the out loop of GBZ $C_{\beta}^{p}$ gives the energy spectra of particles with $\mathrm{Re}(E(\beta_{p})) > 0$.

Figure \ref{gbz}(b) shows the GBZ of critical case between the topologically nontrivial and trivial phases. The two reciprocal loops of GBZ intersect at a critical value $\beta_{c} = - 1$, where the energy gap closes with $E(\beta_{c}) = 0$ as shown in Fig.\ref{gbz}(e). This can be explained as following. The non-Bloch Hamiltonian of the non-Hermitian Kitaev model has an additional symmetry $H^{*}(\beta) = H(\beta^{*})$, which requires an additional condition on the critical values $\beta_{c}$ with $\beta_{c}^{*} = \beta_{c}$. Thus, the gap closing condition in Eq.\ref{gapcondition} becomes $\beta_{c} = \pm 1$. With help of the condition, we can determine the critical chemical potential by $E_{\pm}(\beta_{c})=0$. For positive chemical potential case, the energy gap closes at $\mu_{c} = 2$ with $\beta_{c} = -1$. While, the reciprocal loops of GBZ can intersect at $\beta_{c} = 1$ and the energy gap closes at $\mu_{c} = - 2$ for negative chemical potential.

The $Z_{2}$ skin effect of the non-Hermitian Kitaev model can be well understood by the conception of GBZ. The eigenstates of particle $|\beta_{p}, E \rangle$ and hole $|\beta_{h}, - E \rangle$ satisfy $|\beta_{p}, E\rangle = \sigma_{x} |\beta_{h}, -E\rangle$ with $\beta_{p}=\beta_{h}^{-1}$. Therefore, a right localized particle eigenstate with $|\beta| > 1$ will be mapped to a corresponding left localized hole eigenstate with $|\beta| < 1$. Then, the non-Hermitian Kitaev model exhibits a $Z_{2}$ skin effect, which is protected by the particle-hole symmetry.

To characterize the emergence of the MZMs, we rewrite the non-Bloch Hamiltonian into the Majorana representation.
With the help of GBZ, a $Z_{2}$ non-Bloch topological invariant is given in terms of the Pfaffian of the Majorana representation matrix:
\begin{eqnarray}
\nu = \mathrm{sgn} [\mathrm{Pf} \left(B(0)\right) \mathrm{Pf} \left(B(\pi)\right)] = \pm 1,
\label{invariant}
\end{eqnarray}
here, $\pm 1$ corresponds to topologically trivial and nontrivial. The non-Bloch Majorana representation matrix is $i B(\beta) = U H(\beta) U^{\dagger} = h_{0}(\beta) - h_{y}(\beta) \sigma_{x} - h_{z}(\beta) \sigma_{y}$ with unitary transformation matrix $U$:
\begin{eqnarray}
U= \frac{1}{\sqrt{2}} \left(
  \begin{array}{cc}
    1 & 1   \\
   -i & i   \\
  \end{array}
\right).
\end{eqnarray}
$\beta_{p/h,0}$($\beta_{p/h,\pi}$) are the values of $\beta_{p/h}$ with $\arg \beta_{p/h} = 0$($\arg \beta_{p/h} = \pi$) on the particle/hole loop of GBZ. $B(0)=(B(\beta_{p,0})+B(\beta_{h,0}))/2$ and $B(\pi)=(B(\beta_{p,\pi})+B(\beta_{h,\pi}))/2$ are the averages of all the loops of the GBZ. Then, we can determine that the superconductor is topologically nontrivial with $\nu = -1$ for $\mu \in [-2, 2]$ and trivial with $\nu = 1$ for other region. The robust MZMs exist at the ends of an open chain when the phase is topologically nontrivial. Thus, the emergence of the robust MZMs can be characterized by the $Z_{2}$ non-Bloch topological invariant. A non-Hermitian BBC is well established for the non-Hermitian Kitaev model.

{\it Non-Hermitian $s$-wave superconductor.--} A typical non-Hermitian spinfull $s$-wave superconductor can be described by the Hamiltonian in real space:
\begin{eqnarray}
&~&H = H_{t}+H_{so}+H_{sc}+H_{z} \nonumber\\
&~&H_{t} = \sum_{j,\sigma\in \{\uparrow,\downarrow\}} \left[ - t_{L} c_{j,\sigma}^{\dag}c_{j+1,\sigma} - t_{R} c_{j+1,\sigma}^{\dag}c_{j,\sigma} - \mu c_{j,\sigma}^{\dag}c_{j,\sigma} \right],\nonumber\\
&~&H_{so} = \alpha_{R} \sum_{j} \left[c_{j,\uparrow}^{\dag} c_{j+1,\downarrow}+c_{j+1,\downarrow}^{\dag} c_{j,\uparrow}
-c_{j+1,\uparrow}^{\dag} c_{j,\downarrow}-c_{j,\downarrow}^{\dag} c_{j+1,\uparrow} \right],\nonumber\\
&~&H_{sc} = \sum_{j} \Delta \left[  c_{j,\uparrow}^{\dag} c_{j,\downarrow}^{\dag}- c_{j,\uparrow} c_{j,\downarrow} \right],\nonumber\\
&~&H_{z} = h \sum_{j} \left[c_{j,\uparrow}^{\dag} c_{j,\uparrow} -c_{j,\downarrow}^{\dag} c_{j,\downarrow} \right],
\label{shamiltonian}
\end{eqnarray}
where $c_{j}^{\dag}(c_{j})$ is the creation (annihilation) operator on site $j$, $\alpha_{R}$ and $h$ are real parameters and denote the strength of spin-orbit coupling and Zeeman field, respectively.

\begin{figure}
\includegraphics[width=0.48\textwidth]{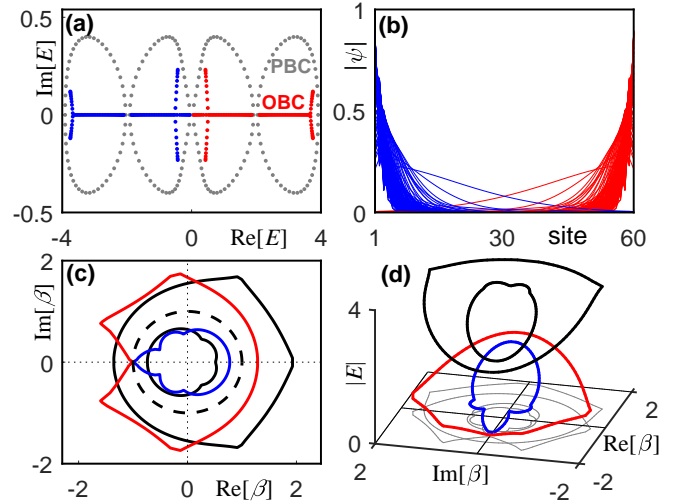}
\caption{(a) The energy spectra in an open $s$-wave non-Hermitian superconductor chain under OBC (red and blue) and PBC (gray). (b) The eigenstates of the open chian. (c) The GBZ (solid curves) and BZ (dashed curves). (d) The OBC energy spectra on the GBZ. The values of the parameters are $t_{L}=1.4$, $t_{R}=0.6$, $\alpha_{R}=0.8$, $\Delta=1.2$, $h = 2$ and $\mu=3.6$.} \label{sgbz}
\end{figure}

Figure \ref{sgbz}(a) shows the OBC energies and PBC energies with $t_{L}=1.4$, $t_{R}=0.6$, $\alpha_{R}=0.8$, $\Delta=1.2$, $h = 2$ and $\mu=3.6$. The four bands OBC energies dramatically collapses comparing with the PBC energies. The eigenstates of particles and holes are localized at the opposite ends of an open chain as shown in Fig.\ref{sgbz}(b).

To understand the non-Hermitian features, the Hamiltonian in Eq.\ref{shamiltonian} is rewritten into a non-Bloch Hamiltonian $H(\beta) = \frac{1}{2} [-\delta (\beta- \beta^{-1}) - (t \beta +  t \beta^{-1} + \mu) \tau_{z} + i \alpha_{R}(\beta-\beta^{-1}) \tau_{z} \sigma_{y}  +  h \tau_{z} \sigma_{z} - \Delta \tau_{y} \sigma_{y}]$, where $\sigma$ and $\tau$ are the Pauli matrices in spin and particle-hole space, respectively. The non-Bloch Hamiltonian preserves the PHS with $\tau_{x} H^{T} (\beta) \tau_{x} = - H (\beta^{-1})$. The GBZ $C_{\beta}$ is given by the trajectories of $\beta_{4}$ and $\beta_{5}$ with $|\beta_{4}| = |\beta_{5}|$ as shown in Fig.\ref{sgbz}(c). The GBZ are two paired reciprocal loops with out two particle loops($|\beta_{p}|>1$) and corresponding inner two hole loops($|\beta_{h}|<1$). Therefore, the superconductor exhibits a $Z_{2}$ skin effect. There are five intersections in the loops of GBZ. The corresponding particle loop(red curve) and hole loop(blue curve) intersect at BZ with $\beta=-1$. As shown in Fig.\ref{sgbz}(d), the energy gap closes only when the corresponding particle and hole loops intersect at $\beta_{c} = - 1$ for $\mu > 0$ and $\beta_{c} = 1$ for $\mu < 0$.

{\it Discussion.--} We established a non-Bloch band theory for 1D non-Hermitian superconductors. Based on the theory, a series of universal physical characteristics have been uncovered, such as two reciprocal loops, $Z_{2}$ skin effect. These results will be crucial for understanding superconductivity under the ubiquitous environment couplings. Our results can be extended to non-Hermitian superconductors with other symmetry such as the Floquet non-Hermitian superconductors. In the presence of periodic driving, the Floquet Hamiltonian also preserves a PHS. The skin modes of the Floquet non-Hermitian superconductors are $Z_{2}$ skin modes. Furthermore, the periodicity in quasienergy will enrich the non-Hermitian features in non-Hermitian superconductors.

{\it Acknowledgements.--} X. Yang thanks to the fruitful discussion with Zhong Wang. This work was supported by the National Natural Science Foundation of China (No.11874314, 12074150).

\bibliography{reference} %title+Hyperlinks

\end{document}